\documentclass[prl,showpacs,amssymb,floatfix,twocolumn]{revtex4}
\usepackage{amsmath}
\usepackage{graphicx}
\usepackage{epsfig}
\usepackage{dcolumn}
\usepackage{bm}
\usepackage{times}

\def\<{\langle}
\def\>{\rangle}
\def\(({\left(}
\def\)){\right)}
\def\[[{\left[}
\def\]]{\right]}

\newcommand{\be}{\begin{equation}}
\newcommand{\ee}{\end{equation}}
\newcommand{\bea}{\begin{eqnarray}}
\newcommand{\eea}{\end{eqnarray}}

\begin{document}

\title{A Non-Disordered Glassy Model with a Tunable Interaction Range}

\author {Frauke Liers$^1$, Enzo Marinari$^2$, Ulrike Pagacz$^1$,
  Federico Ricci-Tersenghi$^2$ and Vera Schmitz$^1$}

\affiliation{$^1$ Institut f\"ur Informatik, 
  Universit\"at zu K\"oln, Pohligstrasse 1, D-50969 K\"oln, Germany.\\
  $^2$ Dipartimento di Fisica, INFM-CNR and INFN - Sezione di Roma 1,
  Universit\`{a} di Roma ``La Sapienza'', P.le A. Moro 2, I-00185
  Roma, Italy}

\begin{abstract}
  We introduce a non-disordered lattice spin model, based on the
  principle of minimizing spin-spin correlations up to a (tunable)
  distance $R$.  The model can be defined in any spatial dimension
  $D$, but already for $D=1$ and small values of $R$ (e.g.\ $R=5$) the
  model shows the properties of a glassy system: deep and well
  separated energy minima, very slow relaxation dynamics, aging and
  non-trivial fluctuation-dissipation ratio.
\end{abstract}
  

\maketitle


In a supercooled liquid the viscosity increases abruptly by several
order of magnitude in a narrow temperature range, and eventually
undergoes a dynamical arrest, that can be observed on any accessible
time scale: this is the essence of a fascinating physical phenomenon,
the so called ``dynamical glass transition''. A vast scientific
literature has been dedicated to its study:
see~\cite{CavagnaReview,KobReview} for interesting reviews on the
subject.  A theoretical understanding of this effect must be based on
a reliable modeling of the underlying material: ideally one would like
to have at hand very simple models that reproduce the main features of
glass-former liquids. From the point of view of a numerical approach,
off-lattice simulations are extremely costly in term of computational
resources: because of that lattice models where each node of the
interacting network contains some degrees of freedom (a binary spin
variable in the simplest case) play an important role.

An analytic approach needs some reasonable approximation: mode
coupling theory \cite{CavagnaReview,KobReview}, for example, helps to
shed some light on the problem.  At the mean-field level, and more
precisely on a fully connected lattice, the solution of the disordered
$p$-spin model~\cite{GM} is now very well understood~\cite{KTW}, and
it turns out to be equivalent to the mode-coupling theory (based on
systems where the Hamiltonian does not include quenched disorder).
The main prediction of these mean-field theories is that below the
dynamical glass transition $T_g$, which is higher than the
thermodynamical critical point $T_k$, the relaxation dynamics is not
able to bring the system to equilibrium in any sub-exponential time
(in the system size). Consequently the system relaxes to the so-called
threshold energy, which is higher than the equilibrium energy, and the
off-equilibrium dynamics shows aging on any measurable time
scale~\footnote{Off-equilibrium simulations can be run for very large
  system sizes, and the exponential time scale is practically
  unreachable.}. During the aging dynamics the fluctuation-dissipation
ratio is different from the one expected in equilibrium, and its
behavior presents peculiar and distinctive features.

The mean field scenario is well characterized and well understood, but
it is still unclear how to adapt it to real systems.  In finite
dimensional systems the lifetime of metastable states is limited:
eventually, during the aging dynamics, a bubble of the equilibrium
state will nucleate and will grow up to the system size.  Nonetheless
the time for nucleating and growing the equilibrium phase may be
extremely large, especially close to a critical point.  Moreover the
Hamiltonian of a real glass-former does not contain any quenched
disorder: this is a crucial difference from the $p$-spin model. The
frustration, which is the main ingredient for the slow relaxation, is
self-induced by the relaxation dynamics: a realistic model of a glassy
system should contain no quenched disorder.

The ``kinematic models''~\cite{RS} are glassy models with no quenched
disorder, where the evolution is governed by a specific dynamical
rule, but it does not correspond to a relaxation on a well-defined
energy landscape: they are interesting but they cannot undergo a true
thermodynamical phase transition, and we will not consider them in the
following.

Not so many ``Hamiltonian glassy models'' without quen\-ched disorder
are available. Shore, Holzer and Sethna~\cite{HSS92} introduced and
analyzed a model with a competition in the interaction, and a tiling
$2D$ model. Newman and Moore~\cite{NM99} have discussed a $2D$ model
with a triangle $3$-spin interaction, that can be solved exactly: it
does not show a sharp glass transition but undergoes a severe slowing
down. Biroli and M\'ezard~\cite{BM02} defined a very simple lattice
glassy model, where the legal positions of the particles are
restricted by hard ``density constraints'': the model is versatile,
since it does not depend on the detailed feature of the underlying
lattice, but its energy landscape is somehow drastic, in the sense that
a configuration is either legal ($E=0$) or forbidden
($E=\infty$). Cavagna, Giardina and Grigera~\cite{CGG03} have
discussed a $2D$ model based again on competing interactions
(respectively with $4$ and $5$ spins). It is clear that enlarging this
collection would be appropriate: some of these interesting models are
indeed strictly two dimensional or depend on the detailed lattice
structure. In some other cases one can observe a very slow domain
growth, but once the time is appropriately rescaled the growth process
does not differ qualitatively from the dynamics of the pure Ising
model.


In this note we define and analyze a new non-disordered glassy model,
that can be defined on any lattice structure, in any dimension, and is
based on a simple physical principle: the minimization of
correlations.  Our model is also, following the route of its mean
field predecessor, a good candidate for providing coding for an
effective and secure communication.

We are inspired from the Bernasconi mean field model~\cite{GO,BE},
where one is interested in finding the assignment to $N$ Ising spins
defined on a linear lattice that minimizes the sum
$\mathcal{H}_\text{B} \equiv \frac{1}{N-1} \sum_{d=1}^{N-1} \left(
\sum_{j=0}^{N-d-1} \sigma_j \sigma_{j+d} \right)^2$ of the squared
spin-spin correlations.  We have used here open boundary conditions,
but the Bernasconi model, like our model, is also interesting when
defined with periodic boundary conditions.  The Bernasconi model has a
very rough energy landscape \cite{BOME,MPR1}, with deep minima
separated by extensive energy barriers, making the search for global
minima (low auto-correlation binary sequences) a very difficult task.
The theoretical analysis of this model predicts a thermodynamical
phase transition with one step of replica symmetry breaking (1RSB), in
the same universality class of the $p$-spin model, preceded by a
dynamical glass transition.  Extensive numerical simulations have
shown that the energy relaxation stops before reaching the ground
state (GS) energy, and the aging regime persist for extremely long
times.  For these reasons the Bernasconi model is the perfect
prototype of a glassy mean field model.

We define our model by adopting the principle of minimizing spin-spin
correlation functions, but using an interaction that is local in
space: in $D$ dimensions the Hamiltonian reads
\begin{equation}
\mathcal{H}_D =
\mathcal{N}_{N,R}
\sum_{\vec{x} \in \Lambda} 
\left[
\sum_{d=1}^{\text{max }d\text{ in }\mathcal{R}(\vec{x})}
\left(
\sum^{\text{couples in }\mathcal{R}(\vec{x})}_{\text{ at distance }d} 
\sigma_j \sigma_k 
\right)^2
\right]
\label{HD}
\end{equation}
where $\Lambda \subset \mathbb{Z}^D$ is a finite volume of cardinality
$N$, $\mathcal{R}(\vec{x})$ is a hypercube of size $R^D$ (or its
intersection with $\Lambda$ if open boundary conditions are used)
centered around site $\vec{x}$, and $\mathcal{N}_{N,R}$ is a
normalization constant that guarantees good $R \to\infty$ and $N \to
\infty$ limits.  The sum over $d$ is for distances going from one up
to the maximum distance contained in $\mathcal{R}(\vec{x})$. One can
also define the model by only considering correlations on the $D$
axis: from the point of view of computational cost this is a far
better choice.

With eq.~(\ref{HD}) we are aiming at minimizing correlations in blocks
of linear size $R$. Since correlations at short distances are
typically the strongest, the overall effect is to have low-energy
configurations showing very weak correlations on all length scales.
This seems to us a very solid first principle to build a glassy model:
glass-formers show no long range order in the two-point correlation
functions and we are somehow enforcing this condition in the
Hamiltonian.  The tunable interaction range $R$ is a novel and very
useful feature of our model.  The $R\to\infty$ limit gives back the
Bernasconi model in $D=1$, and for $D>1$ provides new and potentially
interesting mean-field models. As for the Bernasconi model here we can
get a good basis towards effective coding: the introduction of new,
free parameters ($D$ and $R$) that are unknown to the observer could
be of further help.

Let us look in better detail to the $D=1$ version of the model with open
boundary conditions. The Hamiltonian reads
\begin{equation}
\mathcal{H}_1 = \frac{N}{N-R+1}\sum_{i=0}^{N-R}\frac{1}{R(R-1)}
\sum_{d=1}^{R-1} C(d,i,R)^2 \;,
\label{eq:H1d}
\end{equation}
where $C(d,i,R)\equiv \sum_{j=i}^{i+R-1-d} \sigma_j\sigma_{j+d}$, and
$R$ is the tunable interaction range.  We will show, for example, that
the $R=5$ model presents to a very large extent the glassy
phenomenology we have described above (as opposed, for example, to the
$R=6$ model).  This makes clear that already the $D=1$ case, where a
thermodynamical transition is forbidden, is very interesting for the
physics of structural glasses.

Finding and analyzing low energy configurations is our first task. For
$R=N$ (the Bernasconi model) finding a GS is difficult in practice and
requires a time growing exponentially with $N$. For finite $R$ the
model can be solved in a time of order $O(N 2^{2(R-3)})$ by transfer
matrix methods, using the variables $\tau_i \equiv \sigma_{i-1}
\sigma_{i+1}$.

We have computed GS and first excited states for any $N\le 52$ and $R$
from $3$ to $6$ (for some selected $N$ values we have also analyzed
$R$ values going up to $N$), using exact algorithms.  For a given
couple $N,R$ we have determined all exact GS using a clever exhaustive
enumeration scheme called \emph{branch-and-bound} (b\&b). In b\&b, the
problem is solved using recursion. In a branching step, one of the
variables $\sigma_i$ is chosen, and it is eliminated by creating two
subproblems in one of which $\sigma_i = +1$, and in the other
$\sigma_i = -1$. The latter are solved recursively. A subproblem is
solved by determining upper and lower bounds on its optimum solution
value. The energy value $E^{ub}$ of the best known configurations
serves as upper bound. The latter are updated whenever configurations
with energy $E^{ub}$ or better could be determined. If a subproblem's
lower bound attains a higher value than $E^{ub}$, no configuration
with lower energy can be contained in it. Thus, it can be excluded
from further consideration (fathoming step). For designing a
practically effective algorithm, it is crucial to employ a strong
lower bound with which subproblems can be fathomed early, keeping
their total number reasonably small. For additionally determining all
excited states with value at most $x\%$ away from optimality, we
fathom a node only if its lower bound is worse than $(1.0 +
\frac{x}{100})E^{0}$, being $E^{0}$ the exact GS energy (determined in
an earlier run of b\&b).  Mertens~\cite{Me96} has developed a b\&b
approach for determining exact GS for the model restricted to $R =
N$. He could solve the problem up to $N=60$, which marks the world
record. Following \cite{Me96}, we first narrow the search space by
exploiting the fact that if the configuration $\sigma_1\ldots
\sigma_n$ is a GS also its reversal $\sigma_n\ldots \sigma_1$ is a
GS. The same is true when each odd spin and/or each even spin is
multiplied by $-1$.  We restrict ourselves to representatives of these
symmetry classes.

A lower bound on $\mathcal{H}_1$ is given by minimizing each
$C(d,i,R)^2$ independently. We briefly sketch how we estimate the
value of such a minimum. In a subproblem, several spins might be
fixed, all other spins are yet free. The already fixed parts give some
contribution, say $Z$. We look at all partial sequences of free spins
at distance $d$, i.e., $\sigma_{i+kd},
\sigma_{i+(k+1)d},\ldots,\sigma_{i+ld}$ that are framed by fixed
spins. Depending on whether $Z$ is negative or not, the smallest
possible value of $C(d,i,R)$ is either achieved when the free spins
are all set equal or all alternating. The bound that we calculate from
this is stronger than the one presented in \cite{Me96} in the sense
that we need to enumerate considerably less subproblems. For small and
medium $R$ we get better performance if we start branching by fixing
the spins in the middle of the sequence, expanding towards the
boundaries, than if we start branching on the spins along the
boundary, moving `inwards'.


\begin{figure}[ht]
  \centering 
\includegraphics[width=\columnwidth,angle=0]{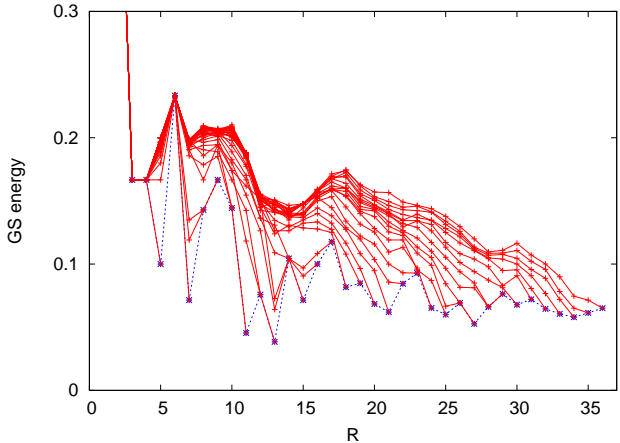}
\caption{GS energy as a function of $R$, for different $N$ values.}
\label{F_ENEGS}
\end{figure}

\begin{figure}[ht]
  \centering 
\includegraphics[width=\columnwidth,angle=0]{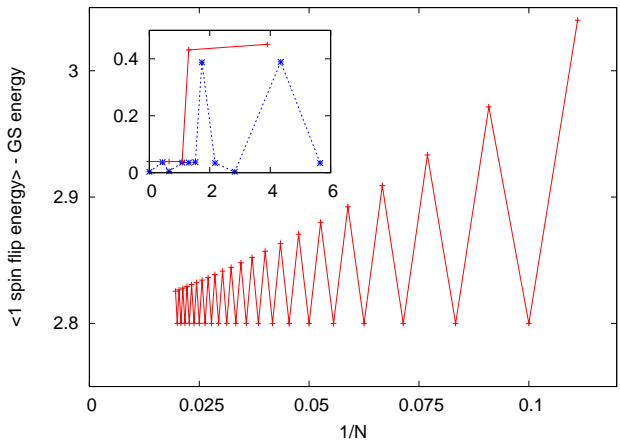}
\caption{R=5. Average energy of states at one spin flip from the
  GS. The energy gap in $\mathcal{H}_1$ is $4$. In the inset the
  probability distribution of these energies for $N=50$ (dashed) and
  $N=51$ (continuous).}
\label{F_1SF}
\end{figure}

\begin{figure}[ht]
  \centering 
\includegraphics[width=\columnwidth,angle=0]{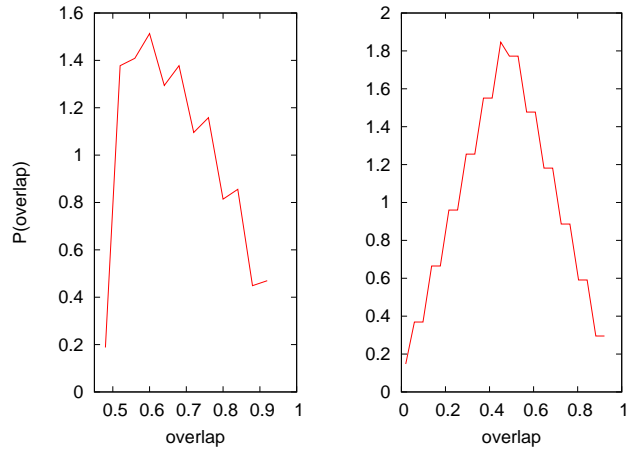}
\caption{Probability distribution of the overlap among the GS
 and the first excited states. $N=50$ (left) and $N=51$
  (right).}
\label{F_OVER01}
\end{figure}

In Fig.~\ref{F_ENEGS} we show the GS energy as a function of
$R$. Different full lines correspond to different $N$ values
(increasing from bottom to top). The lower dashed line joins data
points with $R=N$. For $N \gg R$ (upper lines for low $R$ values) the
data accumulate on a limiting curve: the thermodynamic model with $R$
fixed is well defined.  For $R \le 12$ we worked out an iterative
procedure for computing the GS configurations for any $N$ value.  The
$N \gg R$ limiting curve has, like the infinite range Bernasconi
model, an erratic $R$ dependence, but smoother than in that case.

We have run extensive simulated annealing experiments for $R \le 13$
(to be better discussed below) and we have identified models where the
energy relaxes fast to the GS one, e.g.\ $R=3,6$, and models showing a
very slow relaxation, e.g.\ $R=5,7$.  In the rest of the Letter we
focus on the $R=5$ model, which seems the best compromise between very
short range interactions and glassiness on quite long time
scales~\footnote{We do expect an even more glassy behavior for models
  with larger values of $R$, but MC simulation running time grows like
  $O(R^2)$.}.

In Fig.~\ref{F_1SF} we show the average energy difference $\delta$ of
states at one spin flip from the GS as a function of $N^{-1}$.  The
limit $N\to\infty$ can be estimated very reliably, and is close to
$2.8$, i.e.\ more than ten times the gap (that in these units is equal
to $0.2$).  In the inset we plot the probability distributions of
these energies for $N=50$ and $N=51$: they make clear that the large
mean value of $\delta$ comes from a quasi totality of configurations
that have energies that are much larger than the GS one.  The
distributions shown in the inset only depend, for large $N$, on $N$
being odd (where there is a single GS) or even (where there is a
$N/2+1$-fold degeneracy of the GS).  This is a strong evidence that GS
are surrounded by high barrier of the order of several energy
gaps. This property is shared also by low energy configurations, and
it is what makes the energy landscape somehow golf-course-like.

In Fig.~\ref{F_OVER01} we show the probability of the overlap
$q\equiv\frac1N \sum_i \sigma_i^{(GS)} \sigma_i^{(FIRST)}$ among the
GS and the first excited states, for $N=50$ and $N=51$. We consider
all first excited states and compute $q$ with the GS that is closest
in Hamming distance. The qualitative difference of the two
distributions for $q$ smaller than $1/2$ is connected to the fact that
the GS is not degenerate for odd values of $N$. These probabilities
are peaked at a value of $q$ different from one, and by increasing $N$
the mean value of $q$ stays well below $1$.  The number of first
excited states very similar to the GS is small; they are typically far
from the GS.  This is a further hint towards a glassy nature of the
system.

\begin{figure}[t]
\centering 
\includegraphics[width=\columnwidth]{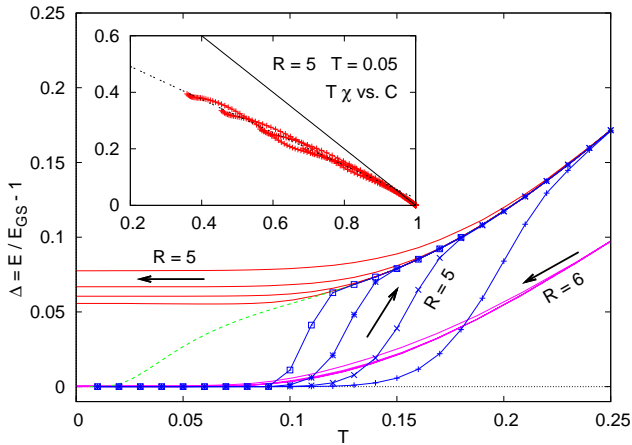}
\caption{Cooling and heating experiments: for $R=6$
  the model fastly relaxes to the GS, while for $R=5$ it has a glassy
  dynamics. The dashed line is the exact energy for $R=5$.
  Inset: integrated response versus correlation ($t_w$ values as in
  Fig.~\ref{F_AGING}).}
\label{F_ANNEALING}
\end{figure}

We discuss now the finite $T$ properties of the model (mainly for
$R=5$).  The system size is always $N=10^6$.  In the main panel of
Fig.~\ref{F_ANNEALING} we show the relative difference $\Delta$
between the system energy and the GS energy during very slow annealing
and heating experiments (marked by leftward and rightward arrows
respectively)~\footnote{We use a linear temperature schedule with
  $\Delta T=0.01$ and run $2\,t$ MCS at each temperature, with
  $t=10^2,10^3,10^4,10^5$}.  Temperature values look small, but this
depends on the factor $1/(R(R-1))$ used in Eq.(\ref{eq:H1d}): the
relevant energy scale is the gap, which is $0.2$ for $R=5$.  While the
$R=6$ model fastly relaxes to the GS energy, the $R=5$ one shows a
very slow relaxation: a tentative extrapolation to the adiabatic limit
using an inverse power of the running time returns $\Delta \simeq
0.048$, i.e.\ is roughly 5\% above the GS energy.  Results from
heating experiments look like a crystal melting, although the
dependence on the heating rate is strong.  For $R=5$ the exact energy
is plotted with a dashed line: it is worth noticing the presence at $T
\simeq 0.038$ of a secondary peak in $C_v$, that seems to enhance
hysteresis in temperature cycles.

\begin{figure}[t]
\centering 
\includegraphics[width=\columnwidth]{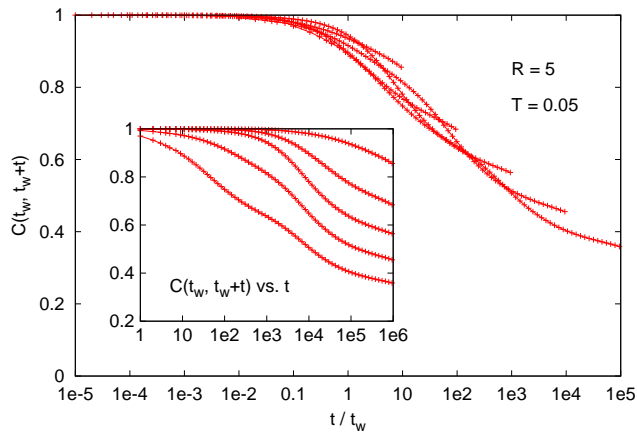}
\caption{Correlation function $C(t_w,t_w+t)$ as a function of $t$
  (inset) and of $t/t_w$ (main panel). Values of $t_w$ are
  $10,10^2,10^3,10^4,10^5$. }
\label{F_AGING}
\end{figure}

We show in Fig.~\ref{F_AGING} the two-time autocorrelation function
$C(t_w,t_w+t) \equiv N^{-1} \sum_i \sigma_i(t_w) \sigma_i(t_w+t)$ for
different values of the waiting time $t_w=10,10^2,10^3,10^4,10^5$.
When plotted as a function of $t$ (inset), the aging behavior is
clear, and very similar to the one observed, for example, in a
Lennard-Jones mixture \cite{KobReview}.  The oscillations are maybe
due to the deterministic nature of the model as in Ref.~\cite{NM99}.
There is a strong dependence of the decay rate on the waiting time
$t_w$.  We show in the main panel how data collapse when plotted
versus $t/t_w$: again, we have a very good agreement with the behavior
observed in glassy systems.

In order to test more quantitatively the out of equilibrium regime, we
have also measured the integrated response to an infinitesimal field
switched on at time $t_w$. We have used the algorithm described in
Ref.~\cite{FedeFDR}. We show in the inset of Fig.~\ref{F_ANNEALING}
the usual plot of the integrated response $\chi(t_w,t_w+t)$ versus the
autocorrelation $C(t_w,t_w+t)$ parametrically in $t$: in the region
where $C$ is not too close to $1$ the data follow a line of slope
smaller than $1$ (in absolute value) as for the $p$-spin model and for
glass-formers.

We have defined a class of models that are potentially good
descriptions of glasses.  We have shown that already one of the
simplest models of our class, the $D=1$ and $R=5$ model, has glassy
properties. We have analyzed the low energy landscape (introducing a
new effective bound in the optimization process), and used Monte Carlo
dynamics to qualify its finite $T$ behavior. There is much interesting
work left, on the mathematical analysis of the model, on the study of
different values of $R$ and the Kac limit, and on the $D>1$ problem.

We acknowledge interesting discussions with A. Billoire and
S. Franz. We are aware that T. Sarlat, A. Billoire, G. Biroli and
J.-P. Bouchaud are studying a local version of the ROM model.


\end{document}